\documentstyle{article}[11pt]\textwidth=160mm\textheight=220mm \oddsidemargin=0pt
\begin{document}

\title{An arbitrated quantum signature scheme}
\author{Guihua Zeng{\footnote{guihuazeng@hotmail.com}}
  and Christoph H. Keitel{\footnote{keitel@physik.uni-freiburg.de}} \\
Theoretische Quantendynamik, Fakult\"at f\"ur Physik, Universit\"at
Freiburg, \\ Hermann-Herder-Strasse 3, D-79104 Freiburg, Germany}

\date{\today}
\maketitle
\begin{abstract}
 The general principle for a quantum signature scheme is proposed
and investigated based on ideas from classical signature schemes and
quantum cryptography.
The suggested algorithm is implemented by a symmetrical quantum key
cryptosystem and Greenberger-Horne-Zeilinger (GHZ) triplet states
and relies on  the availability  of an arbitrator.
We can guarantee the unconditional security of the algorithm, mostly
due to the correlation of the GHZ triplet states and the use of quantum
one-time pads.
\end{abstract}

\newpage

\section{Introduction}
 Quantum cryptography combines quantum theory with classical cryptography. The main goal of this field is to
take advantage of purely quantum effects to provide unconditionally secure information exchange [1], in contrast
in general to classical methods. Those are mostly very secure due to the complexity
of the system employed, however become increasingly vulnerable with more powerful computers and thus improved
means of handling complexity. Many advances have been put forward in quantum cryptography in recent years,
including enhanced insights in the basic theory [2], quantum key management [3,4], quantum secret sharing [5],
quantum authentication [6] and quantum bit commitment [7]. In particular, quantum key distributions
attracted special interest due to technological advances which allow their implementations in laboratory, and
theoretical investigations, which proofed them to be unconditionally secure [4].

An important issue in cryptography is the reliable assignment of a message to its originator. Equally a
certification appears often useful that a particular person has noted or agreed to a message composed by someone
else. Signature schemes are developed classically so far for this purpose as an addition to a message such that
the message can neither be disavowed by the signatory nor can it be forged by the receiver or a possible attacker
[8]. Up to now, conventional (handwritten) and digital approaches have been employed in practical applications.
While conventional signatures can not be transmitted in the electronic network and are vulnerable with respect
to forgery, digital signatures have been used widely and with considerable success in e-commerce.
However,  classical cryptography and thus also classical signature schemes are in general not unconditionally
secure and are in addition difficult to assign to messages in qubit format.

In this paper, we put forward a quantum signature scheme as a method of assigning messages by quantum methods
to its originator or other users. The algorithm takes advantage of the correlation of GHZ states, various
qubit operations and a symmetrical quantum key cryptosystem. It is shown to be unconditionally secure,
i.e. may not be forged or modified in any way by receiver and attacker. In addition it may neither be
disavowed by the signatory nor may it be deniable by the receiver.

The article is arranged as follows. In section 2, we investigate at first the general principles we demand
for a quantum signature scheme which is then proposed and described in detail in section 3.
The proposed scheme includes an initial phase, a signing phase and a verifying phase. In section 4, the
unconditional security of the proposed algorithm is derived and the quantum signature is shown neither to
be disavowable by the signatory nor to be deniable for the receiver. Conclusions are drawn in section 5.


\section{General Requirements}
Before presenting the proposed algorithm, we put forward several aspects to be expected to be fulfilled
for a quantum signature scheme and which have led us to design the quantum signature algorithm to follow.
Similar to classical digital signatures [8] we demand the following signature rules where only the last
is characteristic for quantum signature schemes:

\begin{itemize}
\item No modifications and no forgery: Neither the receiver nor a possible attacker are able to change
the signature or the attached message after completion. The signature may no be reproduced as well.
\item No disavowals: The signatory may not successfully disavow the signature and the signed message.
It need be possible for the receiver to identify the signatory. The receiver may not successfully deny
the receipt of message and signature.
\item Firm assignments: Each message is assigned anew to a signature and may not be separated from it
afterwards.
\item Quantum nature: The signature involves purely quantum mechanical features without a classical analog
and is therefore by nature non reproducible and may not be disavowed or forged.
\end{itemize}

In analogy to conventional and digital signature schemes, a quantum signature algorithm should consist
also of both a signature and a verification algorithm. These algorithms will also have to be prepared by
an initial phase, which initializes or prepares the system parameters and creates the keys.
As usual the signatory, receiver and possible attacker are referred to as Alice, Bob and Oscar, respectively,
where appropriate. We assume the message to be signed to be carried by a string of qubits $|P \rangle $.
The signing algorithm is denoted $QS_{K}$ with key $K$ to be used in the signature phase.
In the verification phase, the resulting signature $|S\rangle$ with $|S\rangle=QS_{K}(|P \rangle)$
can subsequently be verified using a verification algorithm $QV_{K'}$ with key $K'$. Note the keys
$K$ and $K'$ may be the same (symmetrical key cryptosystem) as assumed here or be different (public key
cryptosystem) [8]. Given a pair $(|P \rangle,|S \rangle)$, the verification algorithm when applied is
required to result ``true'' or ``false'' depending on whether the signature is authentic or forged.

A quantum signature scheme may thus be defined as a five-tuple (${\cal P, S, K, Q}_s, {\cal Q}_v$)
with following abbreviations:

\begin{itemize}
\item ${\cal P}$ is a set of possible quantum messages (qubits).
\item ${\cal S}$ is a set of possible signatures. It may consist of qubits or classical bits.
\item ${\cal K}$ is a set of possible keys. It may be a quantum key or a classical key.
\item ${\cal Q}_s$ is a set of possible quantum signature algorithms.
\item ${\cal Q}_v$ is a set of possible quantum verification algorithms.
\end{itemize}

For each key $|K\rangle\in {\cal K}$, there need be a signature algorithm $QS_K\in {\cal Q}_s$ and a corresponding
verification algorithm $QV_{K'}\in {\cal Q}_v$. $QS_K: {\cal P}\rightarrow {\cal S}$ and $QV_{K'}: {\cal
P}\times {\cal S}\rightarrow \{true, false\}$ are functions such that the following equation is satisfied for every
message $|P \rangle \in {\cal P}$ and for every signature $|S \rangle \in {\cal S}$:

\begin{equation}
QV_{K'}(|P \rangle, |S \rangle)=\left \{
\begin{array}{ll}
true\,\,\,  & if\,\,\, |S \rangle =QS_K(|P \rangle)\\ false\,\,\, & if\,\,\,  |S \rangle \neq QS_K(|P \rangle)
\end{array}
\right.
\end{equation}
We emphasize that the signature $|S\rangle$ and the keys may be composed of quantum or classic bits, but
we require the signature and verification algorithms $QS_K$ and $QV_{K'}$ to be of quantum nature.

We recall that signature schemes are generally divided into two categories, the so called {\it true} and
the {\it arbitrated} signature schemes.
The true signatures can be produced and verified independently by the sender and receiver, respectively.
In this category, the signature algorithm is secret but the verification algorithm is public. A judge
may be called only  to settle possible disagreements or disputes. In an arbitrated signature scheme,
however, all communications involve a so-called arbitrator, who authenticates and validates the signed messages.
In this category, both signature algorithm and verification algorithm are secret. In the arbitrated signature
scheme, the arbitrator is required to be trustworthy, because the arbitrator has access to the contents of the
messages and the signatures. While a true signature scheme is in general favorable, arbitrated digital signature
schemes were shown to be applicable and useful, especially with reduced requirements on the trustworthiness of
the arbitrator [10]. In the following, we develop an arbitrated quantum signature scheme based on the requirements
and definitions in this section.

\section{Description of the proposed algorithm}

The proposed algorithm includes three phases: the initial phase, the signature phase and the verification phase.
The scheme involves three partners, the signatory Alice, the receiver Bob and the arbitrator.
In the initial phase, the three communicators entangle themselves via GHZ states and distribute their secret
keys. In the signature phase, Alice prepares and signs her message and obtains an entangled quantum set of
message and signature. In the verification phase, Bob verifies Alice's signature with the arbitrator's help.

\subsection{Initial phase}

This phase generates the keys, sets up the system and distributes the GHZ particles required for our signature scheme.

Step 1. Generation and distribution of keys:
Alice and Bob begin by obtaining their secret keys $K_a,K_b$, where $K_a,K_b$ are employed in the
communications between Alice and arbitrator and between Bob and arbitrator, respectively. These keys may be
obtained by using standard technologies of quantum and classic cryptography. Our keys here are assumed
to be generated via quantum cryptographic methods (see e.g. BB84 or EPR protocols in [3]) because of their
unconditional security.

Step 2. Generation and distribution of GHZ triplet states: Our algorithm relies crucially on the entanglement
of the three involved communicators Alice, Bob and arbitrator. This shall be established here prior each
communication by a distribution of one particle of GHZ triplet states to each of the three. For convenience, we
assume the arbitrator to create and distribute the GHZ particles in our consideration. When the arbitrator
receives Alice's or Bob's application for an arbitrated communication, he is required to create a string
of GHZ triplet states and then to distribute two particles of each GHZ triplet state to each Alice and Bob
and to keep the remaining one for himself for each GHZ state. As a consequence, arbitrator, Alice and Bob are
entangled because they hold one particle of each GHZ triplet state. The GHZ states for a three particle system
involve eight orthonormal triplet states, while in this article, for convenience, we restrict ourselves to the
state

\begin{equation}
|\psi\rangle=\frac{1}{\sqrt{2}}(|000\rangle+|111\rangle).
\end{equation}

We emphasize for above procedures, that step 1 is finished once the system has been set up, and that it is not
necessary to repeat it in later communications. Step 2 is necessary to be redone for every single communication,
the necessity of which becomes clear in the description of the algorithm.

As a practical consideration we add at this stage that GHZ triplet states have been widely studied in quantum
information science [5,11] and in particular have been successfully implemented experimentally [12,13].
With respect to our demands on the GHZ states in step 2, a practical realization may follow the procedure
presented in [14].
Along those lines the arbitrator may generate a short weak light pulse and
then employ an interferometer to split this pulse into two pulses of
smaller, equal amplitude, following each other with fixed phase relation.
The light is then focused into a nonlinear crystal where some of the pump
photons are down-converted into correlated photon pairs. While the first
part of the set-up is located with the arbitrator, the two down-converted
weak photon beams are separated and sent one each to Alice and Bob.
This approach has been successful for the experimental verification of quantum key sharing
[15], such that it should be feasible in principle also for our proposed algorithm.

\subsection {Signing phase}

This phase corresponds to the actual signature algorithm $QS_K$, i.e. to sign the message $|P\rangle$ with
a suitable signature $|S\rangle$. Following steps are required:

Step 1. Alice creates a string of qubits $|P\rangle$ (information qubits) which carry the message  to be
signed. We assume $n$ qubits in the string, such that $|P\rangle$ reads

\begin{equation}
|P\rangle=\{|p_1\rangle, |p_2\rangle, \cdots, |p_n\rangle\}.
\end{equation}
where the symbol $\{\cdots\}$ denotes a set in this article and $|p_i\rangle$ a single qubit in the string
$|P\rangle$. Any qubit $|p_i\rangle (i=1,2,\cdots, n)$ in $|P\rangle$ can be expressed as a superposition of
the two eigenstates $|0\rangle, |1\rangle$, i.e.,

\begin{equation}
|p_i\rangle=\alpha_i|0\rangle+\beta_i|1\rangle,
\label{eq4}
\end{equation}
where $\alpha_i,\beta_i$ are complex numbers with $|\alpha_i|^2+|\beta_i|^2=1$. Using the above equation Eq.
(\ref{eq4}) Alice's information string of qubits can be represented as

\begin{equation}
|P\rangle=\{\alpha_1|0\rangle+\beta_1|1\rangle , \alpha_2|0\rangle+\beta_2|1\rangle , \cdots,
\alpha_n|0\rangle+\beta_n|1\rangle \}.
\end{equation}

Step 2. The aim for Alice in this step is to create a secret string of qubits $|R\rangle$ which involves random
features but also depends clearly on the information string $|P\rangle$. As a first step Alice relates the key
$K_a=\{|K^1_a\rangle, |K^2_a\rangle, \cdots, |K_a^n\rangle\}$ to a sequence of measurement operators ${\cal
M}_{K_a}$, often referred to as measurement basis which we denote
\begin{equation}
{\cal M}_{K_a}=\{{\cal M}^1_{K^1_a}, {\cal M}^2_{K^2_a}, \cdots, {\cal M}^n_{K^n_a} \}.
\end{equation}
The operators $M^i_{K^i_a}$ are defined to arise from the key $|K^i_a\rangle$ for $i \in \{1,2,\cdots, n\}$ via
$M^i_{K^i_a}|K_a^i\rangle=\lambda_i|K_a^i\rangle$.
There is thus a degree of arbitrariness in the definition of those operators with $\lambda_i$ being the
corresponding eigenvalues.
As a simple example, this may, e.g., be carried out for a key $K_a$ consisting of nonorthogonal states $|a\rangle$ and
$|b\rangle$ (see e.g. the B92 protocol in [4]) by choosing two appropriate operators ${\cal O}_a$ and
${\cal O}_b$, where ${\cal O}_a |a\rangle=\lambda_1|a\rangle$ and ${\cal O}_b |b\rangle=\lambda_2|b\rangle$. This way, Alice
may obtain a string of measurement bases ${\cal M}_{K_a}$ consisting of ${\cal O}_a$ and ${\cal O}_b$ by transferring
$|K_a^i\rangle=|a\rangle$ to $M^i_{K^i_a}={\cal O}_a$ and $|K_a^j\rangle=|b\rangle$ to $M^i_{K^i_a}={\cal O}_b$
for $\{i,j\} \in \{1,2,\cdots,n\}$.
Alternatively Alice may use the measurement basis of polarized photons, e.g. as in the BB84 protocol and let the bit ``1''
(or qubit $|\frac{\pi}{4}\rangle$ and $|\frac{3\pi}{2}\rangle$)
correspond to the diagonal measurement basis and ``0'' (or qubit $|0\rangle$ and $|\frac{\pi}{2}\rangle$)
correspond to the rectilinear measurement basis, or vice versa [6].

After the transformation, Alice is required to measure the information string of qubits $|P\rangle$ using ${\cal
M}_{K_a}$ and obtains
\begin{equation}
|R\rangle={\cal M}_{K_a}|P\rangle=\{|r_1\rangle, |r_2\rangle, \cdots, |r_n\rangle\},
\end{equation}
where $|r_i\rangle={\cal M}^i_{K^i_a}|p_i\rangle$ and denotes the i$^{th}$ qubit in the string of $|R\rangle$. Note
the string $|R\rangle$ is secret, is associated with Alice's message and involves both quantum mechanics and Alice's
actions. It will form an essential part of the full signature scheme.

Step 3. Alice entangles each qubit of the information string $|P\rangle$ with one particle each of her
equally long GHZ particle string to become a particle-pair. This may be implemented by applying a joint
measurement on both particles, such as in a quantum logic gate operation [15].
Each combination generates a four-particle entangled state, involving the three GHZ particles and the
information qubit. Using Eqs. (2) and (4) the four-particle entangled state can be described as follows
\begin{equation}
\begin{array}{rl}
|\phi\rangle_i = & |p_i\rangle\otimes|\psi\rangle\\ =&\frac{1}{2}\{|\Psi^+_{12}\rangle_{a}(\alpha_i
|00\rangle_{Ab}+\beta_i |11\rangle_{Ab})\\
          &+|\Psi^-_{12}\rangle_{a}(\alpha_i |00\rangle_{Ab}-\beta_i |11\rangle_{Ab})\\
          &+|\Phi^+_{12}\rangle_{a}(\beta_i |00\rangle_{Ab}+\alpha_i |11\rangle_{Ab})\\
          &+|\Phi^-_{12}\rangle_{a}(\beta_i |00\rangle_{Ab}-\alpha_i |11\rangle_{Ab})\},
\end{array}
\end{equation}
where the subscripts $a,A,b$ correspond, respectively, to Alice, the arbitrator and Bob.
$|\Psi^+_{12}\rangle, |\Psi^-_{12}\rangle,|\Phi^+_{12}\rangle,|\Phi^-_{12}\rangle$ denote the
four Bell states [16].

Step 4.
Alice carries out $n$ Bell measurements, i.e. for each $i \in \{1,\cdots,n\}$ the
state $|\phi\rangle_i$ in Eq. (8) is projected to one of its four summands written on top of each other.
The effect of this measurement is to disentangle Alice's two particles (information qubit and GHZ
particle) to be in one of the four Bell states and to retain the arbitrator's and Bob's corresponding
GHZ particles to be in a two-particle entanglement state as visible in Eq. (8). Thus, Alice obtains the
following set ${\cal M}_a$  of quantum states
\begin{equation}
{\cal M}_a=\{{\cal M}^1_a, {\cal M}^2_a, \cdots, {\cal M}^n_a\},
\end{equation}
where ${\cal M}^i_a$ may be any of the four Bell states in $\{|\Psi^+_{12}\rangle, |\Psi^-_{12}\rangle,
|\Phi^+_{12}\rangle,|\Phi^-_{12}\rangle\}$, in particular is the result arising from her Bell measurement
on state  $|\phi\rangle_i$ in Eq. (8).

Step 5. Alice obtains the quantum signature $|S\rangle$ for the information qubit string $|P\rangle$
by encrypting ${\cal M}_a$ and the secret qubit string $|R\rangle$ by the secret key $K_a$, i.e.
\begin{equation}
|S\rangle=K_a({\cal M}_a,|R\rangle).
\end{equation}

${\cal M}_a$, even though consisting of quantum mechanical Bell states, may be presented by
classical bits, and thus be encrypted by a classical one-time pad. $|R\rangle$ could be encrypted
by the approach known as ``quantum state operation''. Another way would be to transfer ${\cal M}_a$ into a
string of qubits $|{\cal M}_a\rangle$ and then make measurements on both  $|{\cal M}_a\rangle$ and
$|R\rangle$ via ${\cal M}_{K_a}$.

Step 6. Alice sends the string of information qubits $|P\rangle$ followed by the signature $|S\rangle$
to Bob.

We emphasize again that the signature is associated with $|P\rangle$ because $|R\rangle$
was generated via the string of information qubits. We note also at this state already that
Alice's secret key was crucial in preparing the signature such that it appears difficult at
least for Alice to disavow it in the face of the arbitrator or for Bob and attacker to forge
it. In addition we realize that the separation of message and signature by Oscar would not benefit
him or anybody else because the message is valid only with the correct signature and new messages
will be assigned new signatures.
The arbitrator has been hardly involved up to this stage but this will change in the
verification phase to be discussed in the following.

\subsection {Verification phase}

A verification algorithm $QV_K$ is developed here such that Bob is enabled to verify Alice's signature
$|S\rangle$ and consequently judge the authenticity of the information qubits $|P\rangle$. The verification
process in this scheme requires the help of the arbitrator because Bob does not possess Alice's key
which is necessary for the verification of the signature. The verification phase is executed by the
following procedure:

Step 1. Bob measures his string of GHZ particles which, at this stage, are only entangled to the particles
of the arbitrator.
The measurement is performed such that the two possible outcomes are either $|+x\rangle$ or $|-x\rangle$ with
$|+x\rangle=\frac{1}{\sqrt{2}}(|0\rangle+|1\rangle)$ and $|-x\rangle=\frac{1}{\sqrt{2}}(|0\rangle-|1\rangle)$
(referred to as measurements in the $x$ direction).
The sequence of the results of the measurement ${\cal M}_b$ can thus be expressed as
\begin{equation}
{\cal M}_b=\{{\cal M}^1_b, {\cal M}^2_b, \cdots, {\cal M}^n_b\},
\end{equation}
where ${\cal M}^i_b$ is any of two states in $\{|+x\rangle, |-x\rangle\}$. Encrypting ${\cal M}_b, |S\rangle$ and
$|P\rangle$ with the aid of Bob's key $K_b$, he obtains
\begin{equation}
y_b=K_b({\cal M}_b, |S\rangle,|P\rangle).
\end{equation}
Then, Bob sends $y_b$ to the arbitrator.

Step 2. The arbitrator becomes active now and generates a verification parameter $\gamma$ based on the
communication from Bob, which contains information also from Alice.
After receiving $y_b$, the arbitrator decrypts it using $K_b$, and obtains $|S\rangle,|P\rangle,
{\cal M}_b$. Then the arbitrator decrypts $|S\rangle$ using the key $K_a$, which he has since step 1
of the initial phase.
This gives rise to  $|R'\rangle$, which need be compared with $|R\rangle$. With
 $|R'\rangle$, $|P\rangle$ and ${\cal M}_{K_a}$, the arbitrator then creates a parameter $\gamma$ via
\begin{equation}
\gamma=\left\{
\begin{array}{lrr}
1&& if\,\,\, |R'\rangle=|R\rangle={\cal M}_{K_a}|P\rangle\\ 0&& if\,\,\, |R'\rangle\neq |R\rangle={\cal
M}_{K_a}|P\rangle
\end{array}
\right.
\end{equation}

Step 3. The arbitrator measures or evaluates the states of the particles in his string of GHZ
particles. In previous steps, the arbitrator has obtained already Alice's and Bob's measurement results,
${\cal M}_a,{\cal M}_b$, so that he can easily determine his states using Eq. (8). Equally the arbitrator
may choose an appropriate sequence of measurement operators to measure his string of GHZ particles, and
obtains either way ${\cal M}_t=\{{\cal M}^1_t, {\cal M}^2_t, \cdots, {\cal M}^n_t\}$. Note that
${\cal M}^i_t$ may be $|+x\rangle$ or $|-x\rangle$ with the same definitions as in section 1 for Bob.
Encrypting ${\cal M}_a, {\cal M}_b, {\cal M}_t, |S\rangle$ and $\gamma$ via the key $K_b$, the arbitrator
obtains
\begin{equation}
y_{tb}=K_b({\cal M}_a,{\cal M}_b,{\cal M}_t,\gamma, |S\rangle).
\end{equation}
Following completion of this  procedure, the arbitrator sends $y_{tb}$ to Bob.

Step 4. Bob decrypts $y_{tb}$ and obtains ${\cal M}_a,{\cal M}_b,{\cal M}_t, |S\rangle$ and $\gamma$. These
parameters will turn out essential for Bob for the verification of Alice's signature. This will occur in the
two steps to follow, where the first is to eliminate obvious forgeries quickly while the second is more demanding
but allows for full security.

Step 5. Bob undertakes the first verification for Alice's signature $|S\rangle$ via the parameter $\gamma$. If
$\gamma=0$, the signature has obviously been forged and Bob may reject the message $|P\rangle$ immediately. If
$\gamma=1$, Bob goes on for further verification to the next step.

Step 6. The relation $\gamma=1$ merely shows that the secret string of qubits $|R\rangle$ is correct.
However, this does not fully confirm that the signature $|S\rangle$ is correct because the attacker may have
forged the signature by other means (see Eq. (10)). Thus Bob need a further verification. This will have to be
obtained via the initial correlation of the GHZ triplet states. Taking advantage of ${\cal M}_a$ and ${\cal M}_t$
and a further transformation to be detailed later in Eq. (17), Bob evaluates the information string of qubits
$|P'\rangle$. This Bob has to compare with the original information string of qubits $|P\rangle$.
If $|P'\rangle=|P\rangle$, the signature is completely correct and Bob accepts $|P\rangle$, otherwise, he
should reject it. We emphasize that the result $|P'\rangle$ is obtained from a calculation and not a direct
physical measurement, because Bob's particle has already been measured in step 1 of the verification phase.
However, since ${\cal M}_t$ depends on ${\cal M}_b$, the result of the calculation $|P'\rangle$ is equally
influenced by Bob's measurement. This is useful regarding high security because it prevents eavesdropping
via intercepting Bob's GHZ particle as analyzed in [5].

We note that ${\cal M}_a$ and ${\cal M}_t$ are essential for  Bob to obtain  $|P'\rangle$ as obvious
from  Eq. (8). If, e.g., Alice's result is $|\Psi^+_{12}\rangle$ or $|\Psi^-_{12}\rangle$, Bob's density matrix
of the GHZ particle reads
\begin{equation}
\rho_b=|\alpha_i|^2|0\rangle_{bb}\langle 0|+|\beta_i|^2|1\rangle_{bb}\langle 1|,
\end{equation}
while in the remaining two cases $|\Phi^+_{12}\rangle$ and $|\Phi^-_{12}\rangle$, Bob's density matrix of the GHZ
 particle is
\begin{equation}
\tilde \rho_b=|\beta_i|^2|0\rangle_{bb}\langle 0|+|\alpha_i |^2|1\rangle_{bb}\langle 1|.
\end{equation}
Thus even with  Alice's results $\{{\cal M}^i_a\}$, Bob can only obtain partly information of the qubit $|p_i\rangle$
without the knowledge of ${\cal M}_t$. In order to obtain $|p_i\rangle$,
Bob needs thus ${\cal M}_a,{\cal M}_t$ and in addition simultaneously the following transformations [5],
\begin{equation}
\begin{array}{ll}
|\Psi^+_{12}\rangle_a|+x\rangle_A\rightarrow I, & |\Phi^+_{12}\rangle_a|+x\rangle_A\rightarrow \sigma_x,\\
|\Psi^+_{12}\rangle_a|-x\rangle_A\rightarrow \sigma_z, & |\Phi^+_{12}\rangle_a|-x\rangle_A\rightarrow
\sigma_x\sigma_z,\\ |\Psi^-_{12}\rangle_a|+x\rangle_A\rightarrow \sigma_z, &
|\Phi^-_{12}\rangle_a|+x\rangle_A\rightarrow \sigma_x\sigma_z,\\ |\Psi^-_{12}\rangle_a|-x\rangle_A\rightarrow I, &
|\Phi^-_{12}\rangle_a|-x\rangle_A\rightarrow \sigma_x,
\end{array}
\end{equation}
where $\sigma_i, i=x,y,z$ are the Pauli matrices and $I$ is the identity matrix. How this above transformation
need be employed will be explained in the next paragraph with the help of an example.

We assume for example that Alice's result is $|\Psi^+_{12}\rangle$, so that following Eq. (8)
the arbitrator's and Bob's entanglement state must be
\begin{equation}
|\varphi\rangle_{Ab}=\alpha_i|00\rangle+\beta_i|11\rangle.
\end{equation}
It can be rewritten as
\begin{equation}
|\varphi\rangle_{Ab}=\frac{\sqrt{2}}{2}|+x\rangle_A\left (\alpha_i|0\rangle_b+\beta_i|1\rangle_b\right)+
\frac{\sqrt{2}}{2}|-x\rangle_A\left (\alpha_i|0\rangle_b-\beta_i|1\rangle_b\right).
\end{equation}
Obviously, when the arbitrator's result is $|+x\rangle$, above equation shows that Bob's calculated result is
$\alpha_i|0\rangle+\beta_i|1\rangle$, which just equals $|p_i\rangle$. This means that under the transformation
$I$ Bob can calculate the result $|p_i\rangle$. When, however, the arbitrator's result is $|-x\rangle$,
Bob's calculated result is $\alpha_i|0\rangle-\beta_i|1\rangle$. In this case we do not get the original
information qubit in spite of the absence of forgery. Thus a transformation is necessary which is the reason
of Eq. (17). According to Eq. (17) for the arbitrator's result $|-x\rangle$, Bob makes the transformation
$\sigma_z$ on the state $\alpha_i|0\rangle-\beta_i|1\rangle$. Finally Bob obtains the state of $|p_i\rangle$
which is the same as the corresponding state in the original string $|P\rangle$. This is the proof that the
signature was authentic, while if Bob's results after the transformation Eq. (17) had been different to the
corresponding state in the original string $|P\rangle$, there must have been some sort of fraud. This procedure
via Eq. (17) has to be carried out for each state in the information string $|P\rangle$. Alice's signature is
only successfully verified if all $n$ elements are rederived by Bob in the procedure described above in step 6.

We summarize this subsection and emphasize that the verification phase needs the assistance of an
arbitrator. This becomes clear from steps 2 and 3, however the verification itself is completed mainly
by Bob. This reduces the dependence on the arbitrator somewhat. It also saves the resources of the
network system because the complete execution of the verification by the arbitrator is likely to become a
considerable burden on the network system. The parameter $\gamma$ should be useful for a reasonably large
efficiency of the verification procedure. When $\gamma=0$, the received string $|R'\rangle$ differs from
the original secret string of qubits $|R\rangle$, so that the signature is obviously to be rejected
instantaneously. In this case Bob does not need to make further verifications, so that further efforts are
avoided. When $\gamma=1$, the authenticity however is not confirmed yet because the attacker may have forged
the signature $|S\rangle$ by other means. In the practical situation, for example, in which the $K_a$ has
been discovered without Alice's awareness, the parameter $\gamma$ will not be of any help to discover
this happening. In this case Oscar may forge $|R\rangle$ but without ${\cal M}_a$ may not find $|S\rangle$.
In step 6, with the help of the correlation of the GHZ triplet states, Bob would then discover any fraud.

\section {Security Analysis and Discussion}

The security analysis of the quantum signature scheme is different from what we are used to for quantum key
distributions. In the signature scheme, complete security requires, that the signatory can not disavow the
signature, and that the receiver and the attackers have no possibility to obtain the signature or the
signature keys so that they may forge the signature. In the following we will demonstrate that our proposed
algorithm is unconditionally secure.

\subsection{Impossibility of Forgery}

A dishonest Bob or an attacker may seek to forge Alice's signature, to his own benefit. In the two paragraphs
to follow we show that neither Bob nor any attacker may forge the signature or the message.

We begin by assuming that Bob is dishonest and tries to forge Alice's signature. If  successful,
this is beneficial for him because he can change Alice's signature and design a new signature to
a message favorable to him.
This is impossible, however, because the signature key $K_a$ is secretly kept by Alice and the arbitrator.
As a consequence, Bob can not obtain the correct state $|R\rangle$, which is necessary
for the generation of the signature (please see Eq.(10)). Subsequently the parameter $\gamma$ is not
correct, so that this forgery can be noted when the arbitrator is called to settle a dispute between
Alice and Bob.

The attacker is bound to be without success in our algorithm, because the only public parameters are
$|P\rangle,|S\rangle,y_b, y_{tb}$ and they do not offer any information of the secret keys $K_a$ and $K_b$.
Especially, when the communicators encrypt the messages by a one-time pad algorithm which is relatively easy
to be implemented in quantum cryptography, the security is very high. Even if the attacker do somehow get
hold of  Alice's and Bob's keys, a forgery remains still impossible. This is because the attacker has no
access to Alice's measurement results ${\cal M}_a$, which are secret and are involved in generating the
quantum signature $|S\rangle$ (see again Eq. (10)).
The verification condition $|P'\rangle=|P\rangle$ can not be satisfied without the correct ${\cal M}_a$.
Thus, the correlation of the GHZ triplet state avoids forgery by an attacker.

\subsection{Impossibility of Disavowal for the Signatory}

If Alice disavows her signature, it is very easy to discover it, because Alice's key is contained in the signature
$|S\rangle$. Thus, if Alice and Bob are engaged in a dispute because of Alice's disavowal, they just need to
send the signature $|S\rangle$ to the arbitrator. If the signature $|S\rangle$ contains Alice's key $K_a$, this
signature has been carried out by Alice, otherwise, the signature has been forged by Bob or the attacker.
Therefore, the arbitrator is in the position to judge whether Alice has disavowed her signature.

\subsection{Impossibility of Denial for the Receiver}

A conventional and a digital signature scheme is termed undeniable if Bob can not deny his receiving of
Alice's files. This feature is not generally demanded of a signature but it may be useful for many
 practical applications. Our algorithm contains this property, i.e.  Bob can not disavow his receiving of the
 signature $|S\rangle$ and the information qubit string $|P\rangle$. This is essentially impossible because he needs
 the assistance of the arbitrator in the verification process. In addition we can reduce the dependence on the
 arbitrator by small modifications without losing this property of having an undeniable signature scheme.
In the verification procedure, Bob obtains $y_b$ in step 1 and sends it to Alice rather than to the arbitrator as
in the original version. Then Alice obtains the new signature $|\tilde{S}\rangle=K_a({\cal M}_a,|R\rangle,y_b)$
and sends it to the arbitrator. The arbitrator then modifies $y_{ta}$ in step 3 of the verification phase to be
\begin{equation}
\tilde y_{tb}=K_b({\cal M}_a,{\cal M}_b, {\cal M}_t,\gamma, |\tilde{S}\rangle).
\end{equation}
After this change Alice's and Bob's key are included in the signature $|\tilde{S}\rangle$. Then Bob can not
disavow the fact that the received files have come from Alice, i.e., Bob's receipt of the files is undeniable.

\section{Conclusions}

The general principle and all detailed procedures of a quantum signature scheme have been described and
explained. The similarities to the digital signature scheme were pointed out but emphasis was placed on
the description of the quantum methods in the algorithms such as the use of GHZ states.
Our quantum signature scheme includes three phases: the initial phase, the signature phase and the
verification phase. In the initial phase, all keys are prepared and distributed and in particular an
entanglement is established among the communicators including the arbitrator.
In the signature phase, the quantum signature is generated in association with the message and as a function
of various quantum operations, keys, GHZ states and Bell measurements.
The receiver verifies the authenticity of the quantum signature in the verification phase.
Similar to classical arbitrated signature schemes, the verification of the quantum arbitrated signature scheme
also needs the help of the arbitrator. The proposed algorithm
should be practical in small networks (e.g. local rather than wide-spread network systems).
The security analysis showed that the proposed scheme is unconditionally secure and may neither be disavowed
by the signatory nor may it be deniable by the receiver.

\section{Acknowledgements}

This work was supported by the Alexander von Humboldt-Stiftung under grant number IV CHN 1069575 STP
and by Deutsche Forschungsgemeinschaft (Nachwuchsgruppe within SFB 276).

\newpage

\section*{References}
\begin{enumerate}
\item S. Wiesner. Sigact News. vol. 15, 78, (1983);
C. H. Bennett, G. Brassard, S. Breidbart, and S. Wiesner, Advances in Cryptology: Proceedings of Crypto 82, August
1982, Plenum Press, New York, p. 267.
\item B. Schumacher, Phys. Rev. Lett., Vol.80, 5695 (1998).
\item C. H. Bennett, and G. Brassard. Advances in Cryptology: Proceedings of Crypto 84, August 1984,
Springer - Verlag, 475, (1984); A. K. Ekert,  Phys. Rev. Lett., vol. 67, 661, (1991); C. H. Bennett, Phys. Rev.
Lett., vol. 68, 3121, (1992).
\item C. H. Bennett, F. Bessette, G. Brassard, L. Salvail and J. Smolin, J.Cryptology 5, 3 (1992);
W. T. Buttler, R. J. Hughes, P. G. Kwiat, et. al., Phys. Rev. A, vol. 57, 2379 (1998); P. W. Shor, and J.
Preskill, Phys. Rev. Lett. 85, 441 (2000).
\item M. Hillery, V. Buzek, and A. Berthiaume, Phys. Rev. A, vol. 59, 1829 (1999).
\item G. Zeng and W. Zhang, Phys. Rev. A 61, 032303 (2000).
\item A. Kent, Phys. Rev. Lett. 83, 1447 (1999).
\item B. Schneier, Applied Cryptography:protocols, algorithms, and source code in C,
John Wiley \& Sons, Inc., 1994.
\item D. Greenberger, M. A. Horne, and A. Zeilinger, in Bell's Theorem, Quantum theory, and Conceptions of Universe,
edited by M. Kaftos (Kluwer Academic, Dordrecht, 1989); D. Greenberger, M. A. Horne, A. Shimony, and A. Zeilinger,
Am. J. Phys. vol. 58, 1131 (1990).
\item H. Meijer and S. Akl, Advance in cryptography: Proceedings of Crypto'81, Springer - Verlag, 65 (1981).
\item S. Bose, V. Vedral and P.L. Knight, Phys. Rev. A, vol. 57, 822 (1998).
\item W. Tittel, J. Brendel, H. Zbinden, and N. Gisin, Phys. Rev. Lett. 84, 4737 (2000).
\item D. Bouwmeester, J. Pan, M. Daniell, H. Weinfurter, and A. Zeilinger, Phys. Rev. Lett., vol. 82, 1345 (1999).
\item W. Tittel, H. Zbinden, and N. Gisin, Phys. Rev. A, Vol. 63, 042301 (2001).
\item A. Barenco, D. Deutsch, and A. Ekert, Phys. Rev. Lett. 74, 4083 (1995).
\item P. G. Kwiat, K. Mattle, H. Weinfurter, and A. Zeilinger, Phys. Rev. Lett. 75,4337 (1995).
\end{enumerate}
\end{document}